\documentclass[twocolumn,showpacs,preprintnumbers,amsmath,amssymb]{revtex4}

\usepackage{graphicx}
\usepackage{dcolumn}
\usepackage{bm}

\newcommand{\be}{\begin{equation}}
\newcommand{\ee}{\end{equation}}
\newcommand{\ba}{\begin{eqnarray}}
\newcommand{\ea}{\end{eqnarray}}
\newcommand{\ban}{\begin{eqnarray*}}
\newcommand{\ean}{\end{eqnarray*}}
\newcommand{\braket}[2]{\mbox{$ \langle #1 | #2 \rangle $}}

\newcommand{\ket}[1]{\mbox{$ | #1 \rangle $}}
\newcommand{\bra}[1]{\mbox{$ \langle #1 | $}}

\newcommand{\si}{\sigma}
\newcommand{\demi}{\frac{1}{2}}
\newcommand{\compl}{\begin{picture}(8,8)\put(0,0){C}\put(3,0.3){\line(0,1){7}}\end{picture}}

\newcommand{\one}{\leavevmode\hbox{\small1\normalsize\kern-.33em1}}

\begin{document}

\title{Superluminal hidden communication as the underlying mechanism for quantum
correlations: constraining models}

\author{Valerio Scarani}
\email{valerio.scarani@physics.unige.ch}
\author{Nicolas Gisin}%
\email{nicolas.gisin@physics.unige.ch}

\affiliation{Group of Applied Physics, University of Geneva; 20,
rue de l'Ecole-de-M\'edecine, CH-1211 Gen\`eve 4}

\date{\today}

\begin{abstract}
Since Bell's theorem, it is known that quantum correlations cannot
be described by local variables (LV) alone: if one does not want
to abandon classical mechanisms for correlations, a superluminal
form of communication among the particles must be postulated. A
natural question is whether such a postulate would imply the
possibility of superluminal signaling. Here we show that the
assumption of {\em finite-speed} superluminal communication indeed
leads to signaling when no LV are present, and more generally when
only LV derivable from quantum statistics are allowed. When the
most general LV are allowed, we prove in a specific case that the
model can be made again consistent with relativity, but the
question remains open in general.
\end{abstract}


\maketitle

\section{Introduction}

The quest for models underlying quantum mechanics (QM), i.e.
structures out of which QM could "emerge", is an actual topic of
research in the foundations of physics \cite{emerge}. One of the
features that these models are required to recover is the {\em
non-locality} of the correlations of entangled particles. Since
the work of Einstein, Podolski and Rosen in 1935 \cite{epr}, in
fact, it is known that QM predicts correlations between the
outcomes of the measurements of entangled particles, at any
distance. In a classical world, correlations can be due either to
common information available in the preparation, or from the
exchange of a signal. In 1964, John Bell proved \cite{bell} that
common information at the preparation (the so-called {\em local
variables}, LV) cannot reproduce the quantum-mechanical
predictions: if one still wants to think classically, some
additional communication is needed. Such a communication should
propagate faster than light, because the choices of the
measurement settings can be space-like separated events. A natural
question arises then: can one not find a result analog to Bell's
theorem, that would rule out the possibility of {\em superluminal
communication} (SC), thus fully vindicating the non-classical
origin of quantum correlations?

In particular, one may hope to show that a SC among the particles
cannot be "hidden", that is, that any SC-model would break the
no-signaling condition. But this is not true: if the speed of the
SC is allowed to be infinite in a suitable preferred frame (or
preferred foliation), and the amount of transferrable information
is not restricted, one has the most general example of non-local
variables --- actually this is Bohm's view of his own model
\cite{bohm}. Such a model can be made to match any experimental
prediction with no inconsistency: in particular, it can reproduce
QM, in which the no-signaling condition holds.

For other SC-models, however, consistency with the no-signaling
condition becomes an important issue. For instance, the model
proposed by Eberhard, that uses both SC and LV, allows signaling
as demonstrated by the author himself \cite{ebe}; and so does the
Bohm-Bub model \cite{durt}. A SC-model without LV in which the
preferred frame is replaced by several meaningful frames
associated with the experimental devices \cite{suarez} was also
shown to lead to signaling, even in the case of infinite speed
\cite{scagisin}.

In this paper, we consider a large class of SC-models, namely all
those in which the SC is assumed to propagate in the preferred
frame with {\em finite speed} \cite{sca}. Generalizing a previous
result \cite{scagisin}, we prove that the no-signaling condition
can be broken if LV are absent or are restricted to come from a
quantum state, and present a study of the constraints induced by
the no-signaling condition in the presence of the most general LV
\cite{note1}.

\section{The SC-model and the no-signaling constraint}

\begin{figure*}
\includegraphics[width=10cm]{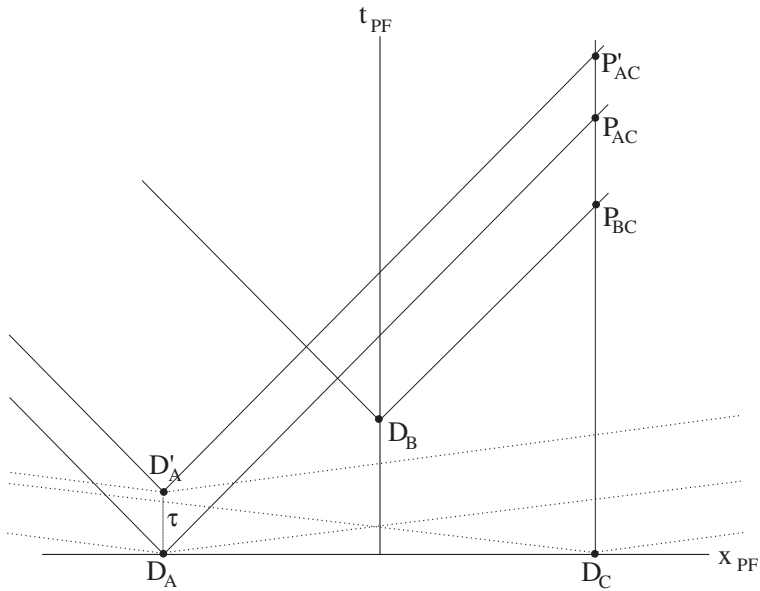}
\caption{Space-time diagram in the coordinates of the preferred
frame (PF) illustrating the non-quantum scenario that leads to
conditions (\ref{cond2}) and (\ref{cond1}). The dots $D$'s are the
detection events, the full diagonal lines are standard light cones
(cones of classical information), the dotted lines are the
"superluminal communication cones". See text for the
explanation.}\label{figxt}
\end{figure*}

As mentioned in the introduction, let's suppose that the "reality"
underlying quantum correlations consists of local variables (LV)
and superluminal communication (SC) with finite speed in a
preferred frame (PF). In such a model, when particle A is measured
before particle B in the PF, so that SC can go from A to B (that
we write A$\rightsquigarrow$B), one recovers the predictions of QM
({\em quantum scenario}); and the same is assumed for the reverse
time-ordering, B$\rightsquigarrow$A. We suppose moreover that all
quantum scenarios are equivalent: as soon as the SC can propagate
from one particle to the other, a given source produces always the
same statistics, compatible with a quantum state $\rho$.

However, when two particles are measured almost simultaneously in
the PF, the SC cannot arrive from one particle to the other
(A$\not\leftrightsquigarrow$B). In particular, if in the quantum
scenario the probabilities are those computed from the singlet
state $\frac{1}{\sqrt{2}}(\ket{01}-\ket{10})$ and can thus violate
Bell's inequality, then, when A$\not\leftrightsquigarrow$B, these
probabilities must be modified in order to become compatible with
LV (no Bell inequality violation). Such a loss of non-locality may
be testable in an experiment, provided the PF is identified and
the sufficient simultaneity is achieved \cite{sca}; however, as
long as only two particles are concerned, the no-signaling
condition does not imply any constraint on the possible models
\cite{scagisin}.

Things are different if we consider {\em three} particles A, B and
C. Let $P(a,b,c|A,B,C)$ be the statistics of a measurement, where
small $x$ are the possible outcomes of the measurement $X$.
Whenever the SC can arrive on each link, e.g.
A$\rightsquigarrow$B, A$\rightsquigarrow$C and
B$\rightsquigarrow$C, the particles give statistics that can be
derived from a quantum state $\rho_{ABC}$ (quantum scenario).
Whenever a link is broken, e.g. when A$\not\leftrightsquigarrow$C,
departures from QM can be expected. In particular, a non-quantum
scenario can be constructed in which:

\begin{itemize}
\item Particles A and C are measured simultaneously in the
preferred frame, hence A$\not\leftrightsquigarrow$C, hence their
correlation can be due only to LV: \ba \mbox{$P(a,c|A,C)$ must
come from LV}\,. \label{cond2} \ea Note that this probability
cannot depend on the choice of the measurement on B, since this
choice may be done later in time.

\item Particle B is measured later, with a time delay sufficient
to ensure A$\rightsquigarrow$B and C$\rightsquigarrow$B, but not
sufficient to ensure communication at the speed of light to arrive
from A or C. It can be shown, see below, that the no-signaling
condition requires \ba \left.
\begin{array}{lcl} P(a,b|A,B,C)&=&P^{QM}(a,b|A,B) \\ P(b,c|A,B,C)&=&P^{QM}(b,c|B,C)
\end{array}\right\} \label{cond1}\ea where $P^{QM}$ are computed
using $\rho_{ABC}$, the state of the source in any quantum
scenario.

\end{itemize}
It is not obvious that conditions (\ref{cond2}) and (\ref{cond1})
are consistent for any choice of the quantum state, and indeed
this will be the main theme of the rest of the paper. Before that,
for completeness let us repeat the construction of the scenario,
already presented in Ref. \cite{scagisin}.

The scenario is depicted in Fig. \ref{figxt}. The particles are at
locations $x_A=-\ell$, $x_B=0$ and $x_C=+\ell$. The dots $D$'s in
the space-time diagram are the detection events. The unprimed
events define the non-quantum scenario: as we said, $D_A$ and
$D_C$ are simultaneous and therefore lie outside the SC-cones
(dotted lines) of each other, whence condition (\ref{cond2}). If A
chooses to delay the measurement by a time $\tau$, so that the
detection of particle A is now $D'_A$, the quantum scenario is
recovered since C$\rightsquigarrow$A, C$\rightsquigarrow$B and
A$\rightsquigarrow$B (follow the SC-cones).

Now, classical information about $D_B$ can arrive at the location
of C at the point labelled by $P_{BC}$: then, $P(b,c|A,B,C)$ can
be estimated. But at that moment, classical information about A
{\em has not yet arrived}, because it will arrive only in $P_{AC}$
or $P'_{AC}$. In particular, the no-signaling condition imposes
that $P(b,c|A,B,C)$ cannot depend either on the measurement done
on A or on whether that measurement was delayed or not. But if the
measurement of A was delayed, we have the quantum scenario, so in
particular $P(b,c|A,B,C)=P^{QM}(b,c|B,C)$ as required in
(\ref{cond2}). The other part of (\ref{cond2}),
$P(a,b|A,B,C)=P^{QM}(a,b|A,B)$ can be derived by the symmetric
argument, supposing that it is C that can delay the measurement
(situation not shown in the figure, for clarity).

\section{The need for local variables}

A first instructive step is taken by supposing that there are no
LV at all, that is, all the correlations are due to SC. In this
case, condition (\ref{cond2}) is replaced by the stronger
condition of independence: \ba
P(a,c|A,B,C)&=&P^{QM}(a|A)P^{QM}(c|C) \label{cond2b}\ea where the
marginals must be those of QM to avoid signaling. Now, it is very
easy to see that this condition and condition (\ref{cond1}) are
incompatible. Consider a source that produces, in the quantum
scenarios, the Greenberger-Horne-Zeilinger state of three qubits
$\ket{GHZ}=\frac{1}{\sqrt{2}}\big(\ket{000}+\ket{111}\big)$, and
suppose that all three measurements are
$A=B=C=\si_z=\ket{0}\bra{0}-\ket{1}\bra{1}$. Then condition
(\ref{cond1}) leads to $P(a=b)=1$ and $P(b=c)=1$; but if $a$ is
always equal to $b$ and $b$ is always equal to $c$, then
$P(a=c)=1$ should hold as well, in contradiction with
(\ref{cond2b}) that predicts $P(a=c)=P(a\neq c)=\demi$. We have
thus proved

{\bf Theorem 1.} {\em In any model of superluminal communication
with finite speed, the assumption that there are no local
variables leads to signaling.}

In some sense, the result of this paragraph is the counterpart of
Bell's theorem for the SC-models that we consider: SC with finite
speed cannot be {\em alone} the cause of quantum correlations,
some LV must be present as well. This was proved in
\cite{scagisin}. In the next paragraph, we extend this result by
showing that a well-defined class of LV model is not enough to
restore the no-signaling condition.

\section{The need for non-quantum statistics}

We can go a step further and require that $P(a,b,c|A,B,C)$ can
always be obtained from a quantum state. This would mean that,
when we arrange a situation in which particles A and C do not
communicate, the statistics are still described by a density
matrix $\tilde{\rho}_{ABC}$ such that the partial state
$\tilde{\rho}_{AC}$ can be described by LV in order to satisfy
(\ref{cond2}). This extension is enough to remove signaling from
the example of the GHZ state described just above: the LV
statistics may be those of the quantum state
$\tilde{\rho}_{ABC}=\demi P_{000}+\demi P_{111}$. However, moving
to other quantum states we can demonstrate the following:

{\bf Theorem 2.} {\em In any model of superluminal communication
with finite speed, the requirement that $P(a,b,c|A,B,C)$ can
always be obtained from a quantum state leads to signaling.}

This follows from a result by Linden and Wootters \cite{linden}
applied to our situation. At least two qubits {\em and one qutrit}
are needed to work out this argument. Consider the state in
$\compl^2\otimes\compl^3\otimes\compl^2$ \ba
\ket{\Psi}&=&\cos\alpha\,\frac{\ket{021}+\ket{120}}{\sqrt{2}}
\,+\, \sin\alpha\,\frac{\ket{000}+\ket{111}}{\sqrt{2}}
\label{psi232}\ea with $0<\alpha<\frac{\pi}{2}$. The statistics of
the sub-systems A-B and B-C are computed from the density matrices
\ba \rho_{AB}=\rho_{CB}&=&\demi\ket{\psi_1}\bra{\psi_1} \,+\,
\demi\ket{\psi_2}\bra{\psi_2}\label{rhos}\ea where
$\ket{\psi_1}=\sin\alpha\ket{00}+\cos\alpha\ket{12}$ and
$\ket{\psi_2}=\sin\alpha\ket{11}+\cos\alpha\ket{02}$. The
statistics of the two qubits A-C is computed from \ba
\rho_{AC}&=&\cos^2\alpha\ket{\Psi^+}\bra{\Psi^+} \,+\,
\frac{\sin^2\alpha}{2}\,\left(P_{00}+P_{11}\right)\ea and violates
the Clauser-Horne-Shimony-Holt (CHSH) inequality for
$\cos^2\alpha>\frac{1}{\sqrt{2}}$ \cite{chsh}. We want to show
that $\ket{\Psi}$ is the only quantum state of A-B-C, pure or
mixed, compatible with the partial traces (\ref{rhos}).

Here is the proof. One starts from $\rho_{AB}$ given by
(\ref{rhos}): since $\ket{\psi_1}$ and $\ket{\psi_2}$ are
orthogonal, any purification of $\rho_{AB}$ can be written \ba
\ket{\Phi}&=&\frac{1}{\sqrt{2}}\big(\ket{\psi_1}_{AB}\ket{E_1}_{CX}
+ \ket{\psi_2}_{AB}\ket{E_2}_{CX}\big) \ea with $X$ an auxiliary
mode and $\braket{E_1}{E_2}=0$. Then, using the Schmidt
decomposition: \ba
\ket{E_1}_{CX}&=&c_0\ket{0}_C\ket{x_{10}}_X+c_1\ket{1}_C\ket{x_{11}}_X\\
\ket{E_2}_{CX}&=&d_0\ket{0}_C\ket{x_{20}}_X+d_1\ket{1}_C\ket{x_{21}}_X\ea
with $\braket{x_{k0}}{x_{k1}}=0$. The rest of the proof goes as
follows: one inserts these expressions into $\ket{\Phi}$, and then
requires that $\rho_{BC}$ is also given by (\ref{rhos}).
Specifically, $\rho_{BC}$ should span a space that is orthogonal
to $\ket{01}_{BC}$ and $\ket{10}_{BC}$. By direct inspection, for
$0<\alpha<\frac{\pi}{2}$, this forces $c_1=d_0=0$, that in turn
implies $c_0=d_1=1$. Using this condition, one can further verify
that $\rho_{BC}$ can be obtained only if
$\braket{x_{10}}{x_{21}}=1$. All in all, this implies means that
\ba\ket{\Phi}_{ABCX}&=& \ket{\Psi}_{ABC}\ket{x}_X\,: \ea
$\ket{\Psi}_{ABC}$ is the only quantum state, pure or mixed,
compatible with the quantum marginals (\ref{rhos}). In particular
then, fixing $\rho_{AB}$ and $\rho_{BC}$ as required by the
no-signaling condition (\ref{cond1}) implies that $P(a,c|A,B,C)$
is the statistics derived from $\rho_{AB}$. For
$\cos^2\alpha>\frac{1}{\sqrt{2}}$, this is non-local, in
contradiction with the spirit of the model (\ref{cond2}).

In conclusion: if, in addition to conditions (\ref{cond2}) and
(\ref{cond1}), we impose that the possible probabilities must
still be describable within quantum physics, then we reach a
contradiction. Thus, if one wants to invoke finite-speed
superluminal communication to describe quantum correlations and,
at the same time, avoid superluminal signaling between observers,
the only hope left lies with local variables distributed according
to {\em non-quantum} statistics.

\section{Most general model}

The additional constraints that we imposed in the previous
sections (no LV, then LV coming from a density matrix) are good
working hypotheses, but rather artificial. If one is ready to
allow a departure from quantum physics by assuming the finiteness
of the "speed of quantum information", then one is also ready to
accept the most general local variable models to describe the
situations where the information is not arrived. Can one still
find a contradiction in this extended framework? That is, are
conditions (\ref{cond2}) and (\ref{cond1}) definitely
contradictory, without any further hypothesis? The answer is, we
don't know. What we do know, is that non-quantum local variables
are enough to remove the contradiction pinpointed in the previous
section, based on the specific state (\ref{psi232}).

To prove this statement, the starting point is to have a
convenient form for the probabilities. Since A and C give binary
outcomes, we can label these outcomes $a,c=\pm 1$. It is easy to
be convinced that any probability distribution of two bits and
another variable (here, the trit $b$) can be written as \ba
P(a,b,c|M)&=&\frac{1}{4}\,\Big[F_M(b)\,+\,a\,A_M(b) \nonumber\\
&&+\,c\,C_M(b)\,+\,ac\,H_M(b)\Big] \label{genprob}\ea where
$M=\{A,B,C\}$ labels the measurements and where the functions
introduced here are submitted to the constraint that all
probabilities must be positive and sum up to one. Note in
particular that $\sum_bF_M(b)=1$. In this notation, the
correlation coefficient A-C is given by \ba E(ac|M)&=&\sum_b
H_M(b)\,.\label{corr}\ea Condition (\ref{cond1}) implies directly
that $F_M$, $A_M$ and $C_M$ must be those that can be computed in
QM, and that the only freedom for an alternative model is left on
$H_M$.

We have to estimate the constraints that are imposed on $E(ac|M)$.
For this, we fix once for all the measurements on the qubits. At
first, we fix also the measurement $B$ and its result $b$. Any
value of $H_M$ is acceptable that satisfies the condition that all
the probabilities are non-negative: \ba
\left.\begin{array}{lcl}P(+,b,+)&=&F_M+A_M+C_M+H_M\geq 0\\
P(-,b,-)&=&F_M-A_M-C_M+H_M\geq 0\end{array}\right\} \label{pp1}\\
\left.\begin{array}{lcl} P(+,b,-)&=&F_M+A_M-C_M-H_M\geq
0\\P(-,b,+)&=&F_M-A_M+C_M-H_M\geq 0
\end{array}\right\}\label{pp2}\ea From (\ref{pp1}) we obtain the
lower bound $H_M\,\geq\,-F_M(b)+|A_M(b)+C_M(b)|\equiv L_M(b)$,
from (\ref{pp2}) the upper bound
$H_M\,\leq\,F_M(b)-|A_M(b)-C_M(b)|\equiv U_M(b)$. In conclusion,
for $B$ and its outcome fixed, all the values of $H$ are possible
that satisfy \ba L_M(b)\,\leq&H_M(b)&\leq\,U_M(b)\,. \ea From this
last equation, using (\ref{corr}), we can immediately derive the
consequent constraint on the A-C correlations: \ba
\sum_bL_M(b)\equiv L_M&\leq E(ac|M) \leq & U_M\equiv
\sum_bU_M(b)\,. \label{bounds}\ea Remember that the source is such
that $E(ac|M)$ violates the CHSH inequality for suitable settings
in the quantum scenario; our goal is to see whether the bounds we
have just derived are tight enough to preserve the violation. Let
$M_{ij}=\{A_i,B,C_j\}$ for $i,j=1,2$: the CHSH inequality reads
$|{\cal{B}}|\leq 2$ where \ba {\cal{B}}&=&
E(ac|M_{11})+E(ac|M_{12}) \nonumber\\&& +E(ac|M_{21})
-E(ac|M_{22})\,. \ea The bounds (\ref{bounds}) impose the
following constraints: \ba {\cal{B}}&\geq& {\cal{L}}\equiv
L_{11}+L_{12}+L_{21}-U_{22}\\ {\cal{B}}&\leq& {\cal{U}}\equiv
U_{11}+U_{12}+U_{21}-L_{22} \ea where $L_{ij}\equiv L_{M_{ij}}$
and $U_{ij}\equiv U_{M_{ij}}$. Thus, the constraints under study
force the violation of CHSH if and only if there exist a family of
four measurements $\{M_{ij}\}$ such that either ${\cal{L}}>2$ or
${\cal{U}}<-2$ holds.

To check this for the state (\ref{psi232}), we recall that the
functions $F_M(b)$, $A_M(b)$ and $C_M(b)$ must be those predicted
by QM. Specifically, let
$\ket{b}=b_0\ket{0}+b_1\ket{1}+b_2\ket{2}$ the eigenstate of
measurement $B$ for the eigenvalue $b$; and the parametrization of
the measurements on the two qubits be given in terms of the
vectors in the Bloch sphere $\hat{n}_X=(\theta_X,\varphi_X)$ for
$X=A,C$. Then we compute $P^{QM}(a,b,c|M)=
\big|\braket{a\hat{n}_A,b,c\hat{n}_C}{\Psi}\big|^2$, write it down
in the form (\ref{genprob}) and thus find \ban
F_M(b)&=&\cos^2\alpha
\,|b_2|^2 + \demi\sin^2\alpha \,(1-|b_2|^2)\,,\\
A_M(b)&=& \demi\sin^2\alpha \cos\theta_A
(|b_0|^2-|b_1|^2)\nonumber\\&&+ \demi\sin 2\alpha
\sin\theta_A\,\mbox{Re}\big[e^{i\varphi_A}(b_0b_2^*+b_2b_1^*)\big]\,,\\
C_M(b)&=& \demi\sin^2\alpha \cos\theta_C
(|b_0|^2-|b_1|^2)\nonumber\\&&+ \demi\sin 2\alpha
\sin\theta_C\,\mbox{Re}\big[e^{i\varphi_C}(b_0b_2^*+b_2b_1^*)\big]
\ean The last step is to maximize ${\cal{L}}$ (respectively
minimize ${\cal{U}}$) over all possible families of four
measurements $\{M_{ij}\}$. This is an optimization over fourteen
real parameters: four for qubit A ($\theta_{A_i}$ and
$\varphi_{A_i}$ for $i=1,2$), as much for qubit C (the analog
ones), and six for the qutrit B, the number of real parameters
needed to define a basis, i.e. an element of $SU(3)$. We
programmed the optimization in Matlab. The result is that
${\cal{L}}$ is always clearly smaller than 2 for any value of
$\alpha$. Specifically, $\bar{{\cal{L}}}=\max_{\{M\}} {\cal{L}}$
starts at $-4$ for $\alpha=0$, then increases to $\sim 0.4$ at the
point $\cos^2\alpha=\frac{1}{\sqrt{2}}$ where the quantum state
$\rho_{AB}$ ceases to violate the CHSH inequality, and finally
reaches exactly 2 for $\alpha=\frac{\pi}{2}$, that is
$\ket{\Psi}=\ket{GHZ}$. As intuitively expected, ${\cal{U}}$
behaves exactly in the symmetric way:
$\bar{{\cal{U}}}=\min_{\{M\}} {\cal{U}}$ starts at 4 for
$\alpha=0$ and decreases down to $-2$ for $\alpha=\frac{\pi}{2}$
\cite{note2}.

Let's summarize: we have studied a state that is entirely
determined by its "quantum marginals" $\rho_{AB}$ and $\rho_{BC}$
if we want to stay within quantum mechanics. However, if we relax
this requirement, several non-quantum functions $H_M(b)$ become
possible --- that quantum probabilities have much built-in
structure is evident e.g. from the fact that $H_M(b)$ must be
bilinear in the vectors $\hat{n}_A$ and $\hat{n}_C$ in the quantum
case, while in the non-quantum case $H_M(b)$ need not even be a
continuous function of these vectors. All this freedom is enough
to break the uniqueness result that holds in the quantum case, so
strongly, that also the non-locality of the marginal distribution
A-C is destroyed. Thus, for the state (\ref{psi232}) that we have
considered and for the CHSH inequality, superluminal communication
with finite speed does not lead to signaling when non-quantum
local variables are allowed. It remains an open problem to
determine whether this conclusion holds in general, whatever the
state and for any possible Bell-type inequality.

\section{Discussion}

We have put constraints on the possibility of using superluminal
communication with finite speed to describe quantum correlations.
Specifically, local variables that yield intrinsically non-quantum
statistics must be provided together with this communication
mechanism, in order to avoid signaling. Whether ultimately such
non-quantum local variables lead to signaling too --- thus ruling
out all models based on finite-speed superluminal communication
--- is still an open question; we sketched a possible
approach to tackle it. The constraints discussed in this paper
should contribute to inspire deeper models for "emergent quantum
mechanics".

\section*{Acknowledgments}

We acknowledge fruitful discussion on this topic with Antonio
Ac\'{\i}n, Lajos Di\'{o}si, Sandu Popescu, Ben Toner and Stefan
Wolf.

\bibliography{apssamp}

\begin{thebibliography}{00}
\bibitem{note1} One might ask why the no-signaling condition should be enforced if we
assume a preferred frame for quantum phenomena. Indeed, there is
no longer a fundamental reason for that requirement. Still,
classical events seem to be conveniently described using special
relativity; and the record of a detection is a classical events
that can be precisely located in space-time.
\bibitem{note2} Obviously, although for some values of
$\alpha$ it happens that $\min_{\{M\}} {\cal{U}}\leq\max_{\{M\}}
{\cal{L}}$, for any given set of measurements ${\cal{U}}\geq
{\cal{L}}$ holds.
\end{thebibliography}

\end{document}